\documentclass[aps,prb,twocolumn,groupedaddress]{revtex4} 

\usepackage{amsmath}
\usepackage{graphicx}
\usepackage{psfrag}
\usepackage{color}
\usepackage{amssymb}

\begin{document}


\newcommand{\tr}{\ensuremath{{\rm tr}}}
\newcommand{\e}{\ensuremath{e}}
\newcommand{\field}[1]{\mathbb{#1}}
\newcommand{\im}{\ensuremath{{\rm Im}}} 
\newcommand{\re}{\ensuremath{{\rm Re}}}
\definecolor{darkgreen}{rgb}{0,0.5,0}

\title{Quantum to Classical Transition of the Charge Relaxation
  Resistance\\ of a Mesoscopic Capacitor}


\author{Simon E.~Nigg}\email[]{simon.nigg@physics.unige.ch}\author{Markus~B\"uttiker}
\affiliation{D\'epartement de Physique Th\'eorique, Universit\'e de
  Gen\`eve, CH-1211 Gen\`eve 4, Switzerland}


\date{\today}

\begin{abstract}
We present an analysis of the effect of dephasing on the single channel charge relaxation resistance of a mesoscopic capacitor in the linear low frequency regime. 
The capacitor consists of a cavity which is via a quantum point contact connected to an electron reservoir and Coulomb coupled to a gate. The capacitor is in a perpendicular high magnetic field such that only one (spin polarized) edge state is (partially) transmitted through the contact. In the coherent limit the charge relaxation resistance for a single channel contact is independent of the transmission probability of the contact and given by half a resistance quantum. The loss of coherence in the conductor is
modeled by attaching to it a fictitious probe, which draws no net current. 
In the incoherent limit one could expect a charge relaxation
resistance that is inversely proportional to the transmission
probability of the quantum point contact. However, such a two terminal
result requires that scattering is between two electron reservoirs
which provide full inelastic relaxation. We find that dephasing of a
single edge state in the cavity is not sufficient to generate an interface resistance. As a consequence the charge 
relaxation resistance is given by the sum of one constant interface resistance and the 
(original) Landauer resistance. The same result is obtained in the high temperature regime due to energy averaging over many occupied states in the cavity. Only for a large number of open dephasing channels, describing spatially homogenous dephasing in the cavity, do we recover the two terminal
resistance, which is inversely proportional to the transmission
probability of the QPC. We compare different dephasing models and discuss the relation of our results to a recent experiment.
\end{abstract}


\maketitle

\section{Introduction\label{sec:introduction}}
Interest in quantum coherent electron transport in the AC regime has
been revived recently thanks to progress made in
controlling and manipulating small high mobility mesoscopic structures
driven by  high frequency periodic voltages at ultra
low temperatures. The state of the art includes the realization of
high frequency
single electron sources, which might be important for
metrology. In Ref.~\onlinecite{Feve:07} this was
achieved by applying large amplitude periodic voltage
pulses of a few hundred MHz on the gate
of a mesoscopic capacitor. The accuracy of this single electron emitter was analyzed theoretically
in Ref.~\onlinecite{Moskalets:07a}. In Ref.~\onlinecite{Kataoka:07}, pulses of surface acoustic waves were
used to transport electrons
one by one on a piezoelectric GaAs
substrate. Two parameter quantized pumping with localized electrical potentials has been demonstrated in Ref.~\onlinecite{Blumenthal:07} and one parameter non-adiabatic quantized charge pumping in Ref.~\onlinecite{Kaestner:07}. These experiments 
use frequencies in the GHz range to control the population and
depopulation of one (or several) localized level(s). 
Thus the dynamics of charge relaxation is of central importance for these experiments.

Of particular interest to us here is the
work of Gabelli et al.~\cite{Gabelli:06}, who succeeded in measuring both the in and out of phase parts of the linear AC conductance
$G(\omega)=I(\omega)/V(\omega)$ of a mesoscopic capacitor at
the driving frequency $\omega\approx 1\, {\rm GHz}$. The capacitor consists of
a sub-micrometer Quantum Dot (QD) connected to an electron reservoir
via a tunable Quantum Point Contact (QPC) and capacitively coupled to a
metallic back or top gate~(see Fig.~1).

The question ``What is the RC-time of a quantum coherent capacitor?''
has been theoretically addressed by B\"uttiker, Thomas and
Pr\^etre~\cite{Buttiker:93b}. In the low frequency regime $\omega \ll 1/\tau_{RC}$,
where $\tau_{RC}$ is the RC-time of the system, the response is
determined by an {\em electrochemical capacitance}
$C_{\mu}$ and a {\em charge relaxation resistance} $R_q$.
\begin{figure}[ht]
{\psfrag{Vqpc}[][][1.2]{$V_{QPC}$}
 \psfrag{1l}[][l][1.2]{$V(\omega)$}
 \psfrag{phi}[l][][1.2]{$V_{\phi}(\omega)$}
 \psfrag{1}{$\color{blue}{N_1}$}
 \psfrag{N}[][r]{$N_{\phi}$}
 \psfrag{Nm1}[][r]{$N_{\phi}-1$}
 \psfrag{Vg}[][][1.2]{$V_g$}
 \psfrag{U}[][][1.2]{$U(\omega)$}
 \psfrag{C}{$C$}
 \psfrag{e}{$\color{red}\varepsilon$}
 \psfrag{T}[][][1.4]{$\color{blue}\mathcal{T}$}
\includegraphics[width=0.4\textwidth]{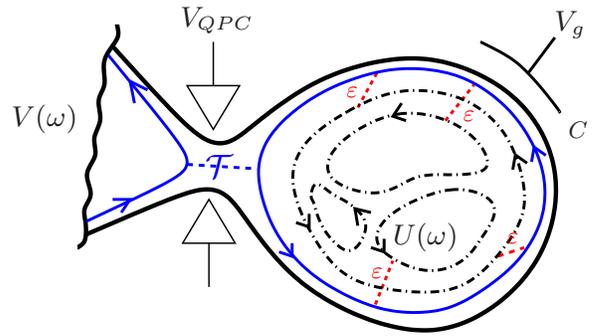}
}
\caption{(Color online) Mesoscopic
Capacitor. The full blue curve represents the
current carrying channel connected to the reservoir via the QPC while
the dashed dotted black curves represent additional localized states, disconnected from the reservoir. As an
example, the
innermost edge state (highest Landau Level) is here split into three localized
states illustrating the possibility of having more than one localized
state per Landau Level. The red dashed lines represent
incoherent scattering between states in the cavity. $\mathcal{T}$ is the
transmission probability of the QPC and $\varepsilon$ is the
inter-channel coupling
strength. $U(\omega)$ is the
Fourier transform of the time-dependent electric potential $U(t)$ inside the
cavity. The functions of the various voltages are discussed in the text.
\label{fig:1a}}
\end{figure}
Together
these determine the RC-time in complete analogy to
the classical case: $\tau_{RC}=R_qC_{\mu}$ . These two quantities
however differ fundamentally from their classical
counterparts. In particular the quantum RC-time obtained from their
product, is sensitive to the quantum coherence of the system and
consequently displays typical mesoscopic fluctuations
\cite{Gopar:96a,Brouwer:97a,Brouwer:97c,Buttiker:07a}. For a system,
with many conducting channels~\cite{Brouwer:97a}, these fluctuations are present
separately in both the capacitance and the resistance. Surprisingly, for a coherent capacitor with a single channel, only the capacitance
fluctuates and the resistance is
found to be constant and given by half a resistance quantum~\cite{Buttiker:93a}
\begin{equation}
R_q = \frac{h}{2e^2} \,.
\label{eq1}
\end{equation}
This quantization has indeed been observed
experimentally~\cite{Gabelli:06} thus establishing a novel
manifestation of quantum coherence in the AC regime.

The claim that the quantization of $R_q$ requires quantum coherence is
perhaps not so astonishing. The interesting question is the length
scale on which coherence is necessary. For the integer quantized Hall
effect~\cite{vKlitzing:80} coherence is necessary only over a cyclotron
radius which is sufficient to establish a Landau level structure. In
fact as discussed in Ref.~\onlinecite{Buttiker:88a} inelastic scattering
(the destruction of long range coherence) can even help to establish
quantization of the Hall resistance. Similarly in quantum point
contacts~\cite{vWees:88,Wharam:88} coherence over the width of the conduction channel is in principle sufficient to establish a step-like structure of the conductance. In contrast, as we will show, the quantization of the charge relaxation resistance requires coherence over the entire capacitance plate (the quantum dot) and not only over the contact region. Thus the quantized charge relaxation resistance in Eq. (\ref{eq1}) is indeed very sensitive to dephasing.   

There is a second important aspect in which the quantized charge relaxation resistance
$R_q=h/(2e^2)$ differs from quantization of a Hall resistance~\cite{vKlitzing:80} or of a ballistic conductance~\cite{vWees:88,Wharam:88}. In both
of these latter cases quantization is associated with perfect
transmission channels which permit unidirectional electron motion
through the sample. In contrast, the quantization of $R_q$ is
independent of the transmission probability of the contact! For a
coherent capacitor plate connected via a single spin polarized channel
to a reservoir the quantization of $R_q$ is truly universal and holds
even in the Coulomb blockade regime~\cite{Nigg:06}.

Of course, no matter how pure the samples are, a spurious interaction of the system
with environmental degrees of freedom, is unavoidable. This
introduces dephasing into the system. 

It is thus of interest to
ask how dephasing affects the quantization of the single channel
charge relaxation resistance and to investigate the crossover from the
coherent to the incoherent regime. Furthermore, in typical
measurements the temperature, though
low compared with the level spacing of the sample, is still
comparable to other relevant energy scales such as the driving
frequency or the coupling strength between cavity and lead. From a theoretical point of view
it is thus desirable to be able to distinguish between thermal averaging and effects due to pure dephasing and to
understand the interplay between these two fundamental
mechanisms. Intuitively, one would expect that in the presence of
strong enough dephasing, the QD starts behaving like an electron
reservoir and thus that a fully incoherent
single channel capacitor should exhibit the two terminal resistance
\begin{equation}
R_q = \frac{h}{e^2} \frac{1}{\mathcal{T}},
\label{eq2}
\end{equation}
where $\mathcal{T}$ is the
transmission probability of the channel through the QPC
connecting the system to the electron reservoir. Interestingly, neither dephasing nor energy averaging (high-temperature limit) lead directly to Eq. (\ref{eq2}). We find that for the QD to become a true electron reservoir it is necessary that many channels participate in the inelastic relaxation process which a true reservoir must provide.

In the present work
we employ a description of dephasing provided by the
voltage and dephasing probe
models~\cite{Buttiker:86,Buttiker:88,Datta:91,DeJong:96,Pilgram:06}, where one
attaches a fictitious probe to the system which can absorb
and re-emit electrons from or into the conductor. If the probe supports
only one channel, we find that the charge relaxation resistance of the {\em fully
incoherent} mesoscopic capacitor is given 
by
\begin{equation}
R_q = \frac{h}{2e^2}+\frac{h}{e^2}\frac{1-\mathcal{T}}{\mathcal{T}}\,,
\end{equation}
Hence, the charge relaxation resistance is given by the sum
of the resistance as found from the (original) Landauer~\cite{Landauer:70} formula 
$h/e^2(1-\mathcal{T})/\mathcal{T}$ and {\em
  one} interface resistance~\cite{Imry:86,Landauer:87} $h/(2e^2)$. Incidentally, as we show below, this is also the
value of $R_q$ obtained in the high temperature limit for the coherent
system, illustrating an interesting relation between single
channel dephasing and temperature induced phase averaging. A hybrid
superconducting-normal conductor provides another geometry with only
one normal narrow-wide interface~\cite{Sols:99}. 

In the next two sections, we introduce the physical system and the
dephasing models. Then in section~\ref{sec:interf-edge-state}, we specialize our model to a
specific form of the scattering matrix appropriate for transport
along edge states of the integer quantum
Hall regime and discuss the main results. Finally, our conclusions
are given in section~\ref{sec:conclusion}.

\section{The mesoscopic capacitor\label{sec:mesoscopic-capacitor}}
The system we consider can be viewed as the mesoscopic equivalent of the ubiquitous
classical series RC circuit. One of the macroscopic ``plates'' of the classical
capacitor is replaced by a QD and the role of the
resistor is played by a QPC connecting this QD to an
electron reservoir. This system is represented schematically in Fig.~\ref{fig:1a}. The curves with arrows represent the transport
channels of the system corresponding physically to edge states of the
integer (spin polarized) quantum
Hall regime, in which the experiment of Ref.~\onlinecite{Gabelli:06} was
performed. By varying the gate voltage $V_{QPC}$, one
changes both the transparency of the QPC and the electrostatic
potential in the cavity. In the present work we take the gate voltage
$V_g$, applied to the macroscopic ``plate'' of the capacitor, as a
fixed voltage reference and
 set it to zero. A sinusoidal AC voltage $V(\omega)$, applied to the
electron reservoir, drives an AC current through the system. 

The low frequency linear AC response of the mesoscopic
capacitor can be characterized~\cite{Buttiker:93b} by an electrochemical capacitance
$C_{\mu}$ and a charge relaxation resistance $R_q$, defined
via the AC conductance as
\begin{equation}\label{eq:2}
G(\omega) = -i\omega C_{\mu}+\omega^2C_{\mu}^2R_q+O(\omega^3)\,.
\end{equation}
The linear low frequency regime is given by $eV_{ac}\ll \hbar\omega\ll\Delta$, where
$V_{ac}$ is the amplitude of the AC voltage and $\Delta$ is the mean
level spacing in the QD.

Even in very clean samples some coupling of
the current carrying edge channel to some environmental states is
unavoidable. For example, we can expect that an electron
entering the QD in the current carrying edge channel (full blue curve
in Fig.~\ref{fig:1a}) may be scattered (red dashed lines in
Fig.~\ref{fig:1a}) by phonons or other electrons into
localized states belonging to other (higher) Landau levels not directly connected to the lead, before being scattered back into the open edge
channel and returning to the electron reservoir. If on the one hand,
this inter-edge state scattering is purely {\em elastic}, the presence
of these closed states is known to lead
to a periodic modulation of the conductance as a function
of gate voltage, the period of which is proportional to the
number of closed states~\cite{Staring:92,Heinzel:94}. Such modulations, with a period corresponding
to about $10$ to $15$ closed states, have indeed been observed
in the experiment of Ref.~\onlinecite{Gabelli:06} at low temperatures for a magnetic field strength of
$1.3\,{\rm T}$. If on the other hand the scattering is {\em inelastic}, such
processes will in general be incoherent, i.e. they will destroy the
information carried by the phase of the electronic wave and hence lead
to dephasing.

The idea of
the present work is to mimic the latter processes using the voltage and
dephasing probe models as illustrated in Fig.~\ref{fig:1b}. The
extension of these models to the AC regime is presented in the next
section. For
simplicity, we will here neglect the contribution of the elastic
processes and focus solely on the inelastic ones.

\section{Voltage and dephasing probe models in the AC
  regime\label{sec:volt-deph-probe}}

\begin{figure}[ht]
{\psfrag{VQPC}[][][1.2]{$V_{QPC}$}
 \psfrag{V}[][l][1.2]{$V(\omega)$}
 \psfrag{phi}[l][][1.2]{$V_{\phi}(\omega)$}
 \psfrag{1}{$\color{blue}{N_1}$}
 \psfrag{N}[][r]{$N_{\phi}$}
 \psfrag{Nm1}[][r]{$N_{\phi}-1$}
 \psfrag{Vg}[][][1.2]{$V_g$}
 \psfrag{U}[l][l][1]{$U(\omega)$}
 \psfrag{C}{$C$}
 \psfrag{DPVP}[][]{DP, VP}
 \psfrag{e}{$\color{red}\varepsilon$}
 \psfrag{T}[][][0.9]{$\color{blue}\mathcal{T}$}
 \psfrag{Ip}[][][1]{$I_{\phi}=0$}
 \psfrag{DPVP}[][]{DP or VP, $I_{\phi}=0$}
\includegraphics[width=0.38\textwidth]{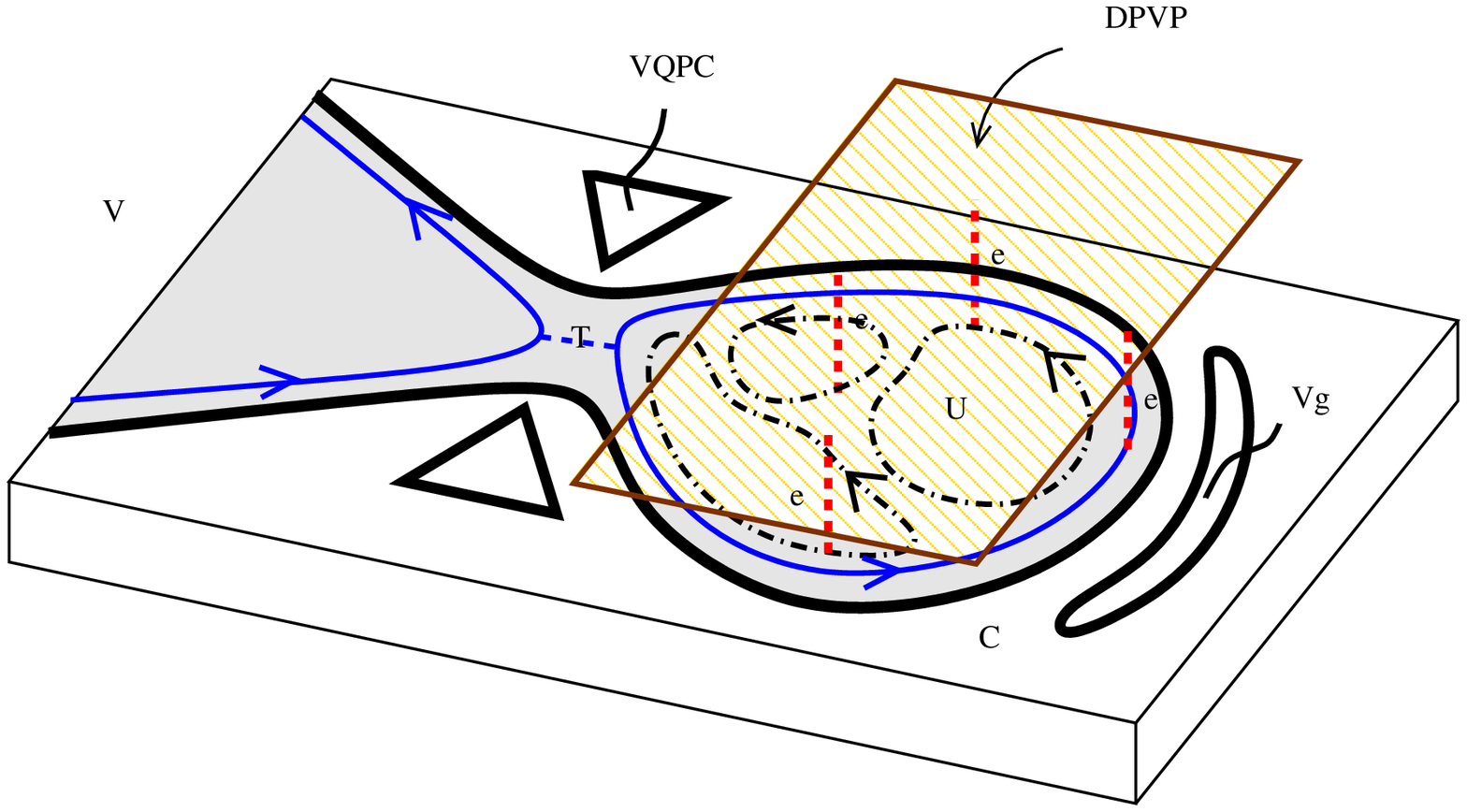}}
\caption{(Color online) 
Schematic representation of the voltage and dephasing probe models. The
incoherent inter-channel coupling depicted in Fig.~\ref{fig:1a}, is mediated by a Voltage Probe (VP)
or a Dephasing Probe (DP) represented as a shaded plane. The entire system including the fictitious
probe, which draws no net current, is again described as a {\em
  coherent} multi-terminal scatterer.
\label{fig:1b}}
\end{figure}

To simulate the loss of phase coherence of electrons inside the
cavity, we attach to the quantum dot a fictitious probe~\cite{Buttiker:86,Buttiker:88,Datta:91,DeJong:96}, which draws
no net current. An electron
entering this probe is immediately replaced by an electron re-injected incoherently into the
conductor. The main advantage of this approach is that the entire
system consisting of the conductor and the probe can be treated as a coherent multi-terminal conductor within
the scattering matrix approach. Some recent applications of this
approach include investigations on the effect of dephasing on quantum pumping
\cite{Moskalets:01,Chung:07}, on quantum limited detection
\cite{Clerk:04} and on photon assisted shot noise
\cite{Polianski:05}. The effect of dephasing on shot noise and higher
moments (counting statistics) has been investigated
in Refs.~\onlinecite{Pilgram:06} and \onlinecite{Forster:07}. A probe which dephases spin states
has been introduced in Ref.~\onlinecite{Michaelis:06}. 

In terms of the spectral current density $i_{\alpha}(E,\omega)$, the
current at the driving frequency $\omega$ into probe $\alpha$ is expressed as
\begin{equation}\label{eq:3}
I_{\alpha}(\omega) = \int dE\, i_{\alpha}(E,\omega)\,.
\end{equation}
The gauge invariant spectral current in turn is given by
\begin{equation}\label{eq:4}
i_{\alpha}(E,\omega) = \sum_{\beta}g_{\alpha\beta}(E,\omega)(V_{\beta}(\omega)-U(\omega))\,,
\end{equation}
where
\begin{equation}\label{eq:5}
g_{\alpha\beta}(E,\omega) =\frac{e^2}{h}F_{\beta}(E,\omega)\tr[\openone_{\alpha}\delta_{\alpha\beta}-S^{\dagger}_{\alpha\beta}(E)S_{\alpha\beta}(E+\hbar\omega)]
\end{equation}
is the (unscreened) spectral AC conductance from probe $\beta$ to
probe $\alpha$ and
$F_{\beta}(E,\omega)=[f_{\beta}(E)-f_{\beta}(E+\hbar\omega)]/\hbar\omega$,
$f_{\beta}$ being the electron distribution function in probe
$\beta$. $V_{\beta}(\omega)$ is the voltage applied to reservoir
$\beta$ and $U(\omega)$ is the Fourier transform of the electric
potential inside the QD, which is
assumed to be homogeneous. The inclusion of this potential, which
accounts for the screening interaction between charges on the
conductor and charges on the gate electrode, is essential
to ensure gauge invariance in the dynamical regime~\cite{Buttiker:93b}. Finally, $S_{\alpha\beta}(E)$ is the
$(N_{\alpha}+N_{\beta})\times (N_{\alpha}+N_{\beta})$ scattering
matrix for electrons with energy $E$ scattered
from the $N_{\beta}$ channels of probe $\beta$ to the $N_{\alpha}$
channels of probe $\alpha$.

In the following we will be interested in the situation where only one
current carrying channel (full blue curve in Fig.~\ref{fig:1b}) connects the QD to the electron reservoir
($N_{1} = 1$),
while the number of channels coupling to the fictitious probe
$N_{\phi}$ is arbitrary.

For the voltage probe, we require that the current into the fictitious
probe vanishes at each instant of time or equivalently at all
frequencies, i.e. $I_{\phi}(\omega)=0$. For the dephasing probe, we require in addition that
the current into the probe vanishes in any infinitesimal energy
interval $dE$ and thus that the spectral current
$i_{\phi}(E,\omega)=0$. This latter condition simulates quasi-elastic
scattering where the energy exchanged is small compared to all other
energy scales. Clearly, with these definitions, a dephasing
probe is also a voltage probe but a voltage probe need not be a
dephasing probe. In both cases however, current conservation implies
that $I(\omega)\equiv I_1(\omega) =
-i\omega CU(\omega)$, where $C$ is the geometrical capacitance of the QD. This relation, together with Eqs.~(\ref{eq:3})
and (\ref{eq:4}), allows us to self-consistently eliminate the
internal potential $U(\omega)$
in the usual fashion~\cite{Buttiker:93b}.
\subsection{Voltage probe\label{sec:voltage-probe}}
From the condition $I_{\phi}(\omega) =0$, we find the AC conductance
\begin{equation}\label{eq:6}
G(\omega) =\frac{I(\omega)}{V(\omega)}=\frac{-i\omega C\chi(\omega)}{\chi(\omega)-i\omega C}\,,
\end{equation}
where
\begin{equation}\label{eq:7}
\chi(\omega) = g_{11}(\omega)-\frac{g_{1\phi}(\omega)g_{\phi 1}(\omega)}{g_{\phi\phi}(\omega)}\,.
\end{equation}
Here and for all of the following, we have introduced the notation
$g_{\alpha\beta}(\omega)=\int dE g_{\alpha\beta}(E,\omega)$. Upon expanding to second order in $\omega$ and comparing coefficients with (\ref{eq:2}),
we find
\begin{equation}\label{eq:16}
C_{\mu} = \frac{C\chi_1}{-iC+\chi_1}\quad\text{and}\quad R_q = -\frac{\chi_2}{{\chi_1}^2}\,,
\end{equation}
with
\begin{equation}\label{eq:14}
\chi_1 =
\sum_{\alpha,\beta}g^{1}_{\alpha\beta}\,\,\, \text{and}\,\,\,\chi_2 =
\sum_{\alpha\beta}\Big(g^{2}_{\alpha\beta}-g^{1}_{\alpha\phi}g^{1}_{\phi\beta}/g^{0}_{\phi\phi}\Big)\,,
\end{equation}
where
$g_{\alpha\beta}(\omega)=g^{0}_{\alpha\beta}+g^{1}_{\alpha\beta}\omega+g^{2}_{\alpha\beta}\omega^2+O(\omega^3)$
and
$\chi(\omega)=\chi_1\omega+\chi_2\omega^2+O(\omega^3)$.
The conductance expansion
coefficients are given in terms of the scattering matrix and its
energy derivatives as
\begin{equation}\label{eq:19}
g_{\alpha\beta}^i = \int dE
f_{\beta}'(E)A_{\alpha\beta}^i(E)\,,\quad (i=1,2,3)
\end{equation}
with
\begin{align}\label{eq:9}
A_{\alpha\beta}^0&=-\frac{e^2}{h}\tr[\openone_{\alpha}\delta_{\alpha\beta}-S^{\dagger}_{\alpha\beta}S_{\alpha\beta}]\,,\nonumber\\
A_{\alpha\beta}^1&=\frac{e^2}{4\pi}\tr[S_{\alpha\beta}^{\dagger}S_{\alpha\beta}'-{(S_{\alpha\beta}')^{\dagger}}S_{\alpha\beta}]\,,\\
A_{\alpha\beta}^2&=-\frac{e^2h}{8\pi^2}\tr[{{S_{\alpha\beta}'}^{\dagger}}S_{\alpha\beta}'-\frac{1}{3}\left(S_{\alpha\beta}^{\dagger}S_{\alpha\beta}\right)'']\,,\nonumber
\end{align}
where $'$ denotes differentiation with respect to $E$ and for compactness we
have suppressed the energy arguments.
In the voltage probe model the electrons in the fictitious lead are
allowed to relax towards equilibrium arbitrarily fast and we thus have $f_{\phi}(E)=f_{1}(E)=1/[1+\exp(\beta(E-E_F))]\equiv f(E)$.

\subsection{Dephasing probe\label{sec:dephasing-probe}}
In contrast to the voltage probe, the distribution function $f_{\phi}(E)$ of the dephasing probe is a priori not
known. The requirement $i_{\phi}(E,\omega)=0$, together with Eq.~(\ref{eq:4}) yields
\begin{equation}\label{eq:8}
G(\omega)=\frac{-i\omega
  C\tilde\chi(\omega)}{\tilde\chi(\omega)-i\omega C}\,,
\end{equation}
where 
\begin{equation}\label{eq:11}
\tilde\chi(\omega)\equiv g_{11}(\omega)-\int
dE\frac{g_{1\phi}(E,\omega)g_{\phi 1}(E,\omega)}{g_{\phi\phi}(E,\omega)}\,.
\end{equation}
The electrochemical capacitance and the charge relaxation resistance
are given in terms of the first and second order frequency expansion coefficients $\tilde\chi_1$ and
$\tilde\chi_2$ as
\begin{equation}\label{eq:12}
C_{\mu}=\frac{C\tilde\chi_1}{-iC+\tilde\chi_1}\quad\text{and}\quad R_q
= -\frac{\tilde\chi_2}{{\tilde\chi_1}^{\ 2}}\,.
\end{equation}
Making use of the unitarity of the scattering matrix, we find explicitly
\begin{equation}
\tilde\chi_1 = \sum_{\alpha\beta}\int dE f'(E)A_{\alpha\beta}^1(E)=\chi_1\,,
\end{equation}
and
\begin{equation}\label{eq:20}
\tilde\chi_2 = \sum_{\alpha\beta}\left(g_{\alpha\beta}^2
  -\int dEf'(E)\frac{A_{\alpha\phi}^1(E)A_{\phi\beta}^1(E)}{A_{\phi\phi}^0(E)}\right)\,.
\end{equation}

Comparing with Eqs.~(\ref{eq:14}), we see that at zero temperature, voltage and dephasing probes
equally affect the AC conductance. At finite temperature, the electrochemical capacitance of the
mesoscopic capacitor does still not distinguish between dephasing and
voltage probes, while the charge relaxation resistance is in principle
sensitive to whether the dephasing mechanism is quasi-elastic
(dephasing probe) or inelastic (voltage probe).

\section{Interfering edge state model\label{sec:interf-edge-state}}
\subsection{Scattering matrix for independent channels\label{sec:scatt-matr-indep}}
We next apply the two dephasing models described in the previous
section to a model for the scattering matrix of the mesoscopic
capacitor in the integer quantum Hall regime introduced in
Refs.~\onlinecite{Gabelli:06} and \onlinecite{Pretre:96},
which is here extended to include a
voltage (dephasing) probe. The special form of
the scattering matrix arises
due to multiple reflections of the electronic wavefunction within the
cavity in close analogy with a Fabry-Perot interferometer.

\begin{figure}
{\psfrag{S1}{$S_1$}
 \psfrag{S}{$S$}
 \psfrag{N}{$N_{\phi}$}
 \psfrag{1}{$1$}
 \psfrag{phi1}[][][0.7]{$\phi_1$}
 \psfrag{phi2}[][][0.7]{$\phi_2$}
 \psfrag{phi3}[][][0.7]{$\phi_3$}
 \psfrag{phi4}[][][0.7]{$\phi_4\dots$}
 \psfrag{S2}{$S_{\varepsilon}$}
\includegraphics[width=0.4\textwidth]{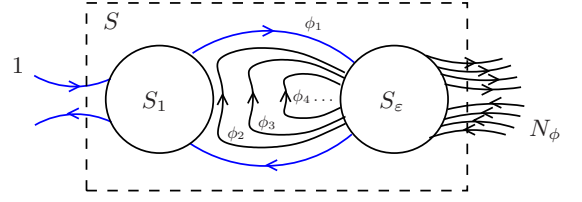}}
\caption{(Color online) $S_1$ is the $(N_{\phi}+1)\times
  (N_{\phi}+1)$ scattering matrices of the QPC
  relating the incoming channel to the $N_{\phi}$ channels inside the
  cavity and $S_{\varepsilon}$ is the $2N_{\phi}\times 2N_{\phi}$
  scattering matrix relating the $N_{\phi}$ channels in the cavity
  with the $N_{\phi}$ channels in the probe. From these two matrices
  one can derive $S$, the total
  scattering matrix relating the incoming channel to the $N_{\phi}$ channels
  in the probe. An electron in the $i$-th channel accumulates a phase
  $\phi_i$ inside the cavity.\label{fig:2}}
\end{figure}

The additional probe, with $N_{\phi}$ channels is coupled to the
single edge channel propagating through the QPC. Clearly $N_{\phi}-1$ channels of the probe
are perfectly reflected at the QPC from within the cavity as depicted
in Figs.~\ref{fig:1b} and \ref{fig:2}. For
simplicity, we shall assume the channels to be independent, which
means that we consider the physical edge channels to coincide with the
eigen-channels of the transmission matrix. Furthermore, we consider a
symmetric QPC and assume that each channel couples identically to the
fictitious probe with strength $\varepsilon$. Then the $(N_{\phi}+1)\times(N_{\phi}+1)$ scattering
matrix $S_1$ of
the QPC and the $2N_{\phi}\times 2N_{\phi}$ scattering matrix $S_{\varepsilon}$
of the fictitious probe have
block diagonal form and may be parameterized as follows
\begin{equation}
S_1=\begin{pmatrix}r_1&t'_1\\t_1&r'_1\end{pmatrix}\quad\text{and}\quad S_{\varepsilon}=\begin{pmatrix}r_{\varepsilon}&t'_{\varepsilon}\\t_{\varepsilon}&r'_{\varepsilon}\end{pmatrix}
\end{equation}
with $r_1=ir$ and
$r'_1={\rm diag}(ir\e^{i\phi_1(E)},\e^{i\phi_2(E)},\dots$ $\dots,\e^{i\phi_{N_{\phi}}(E)})$
where we take $r$ to be real and $\phi_l(E)$ is the phase accumulated by an
electron during one round trip along the $l$-th edge state through the
QD. $t_1=(\sqrt{1-r^2},0,\dots,0)^T$ and
$t^{\prime}_1=(\sqrt{1-r^2}\e^{i\phi_1(E)},0,\dots,0)$.  Finally
$r^{(\prime)}_{\varepsilon}={\rm
  diag}(i\sqrt{1-\varepsilon},\dots,i\sqrt{1-\varepsilon})$ and $t^{(\prime)}_{\varepsilon}={\rm
  diag}(\sqrt{\varepsilon},\dots,\sqrt{\varepsilon})$. 

The total $(N_{\phi}+1)\times (N_{\phi}+1)$ scattering matrix, which
is obtained from the series combination of the two scattering matrices $S_1$ and
$S_{\varepsilon}$, takes the form
\begin{equation}\label{eq:13}
S=\begin{pmatrix}S_{11}&S_{1\phi}\\S_{\phi1}&S_{\phi\phi}\end{pmatrix}\,,
\end{equation}
with
\begin{align}\label{eq:15}
S_{11}&=\frac{i(r+\e^{i\phi_1}\sqrt{1-\varepsilon})}{1+r\e^{i\phi_1}\sqrt{1-\varepsilon}}\,,\nonumber\\
S_{1\phi}&=(\frac{\sqrt{\varepsilon(1-r^2)}\e^{i\phi_1}}{1+r\e^{i\phi_1}\sqrt{1-\varepsilon}},\overbrace{0,\dots,0}^{N_{\phi}-1})\,,\\
S_{\phi 1}&=(\frac{\sqrt{\varepsilon(1-r^2)}}{1+r\e^{i\phi_1}\sqrt{1-\varepsilon}},\overbrace{0,\dots,0}^{N_{\phi}-1})^T\,,\nonumber\\
S_{\phi\phi}&={\rm
  diag}(\frac{i(\sqrt{1-\varepsilon}+r\e^{i\phi_1})}{1+r\e^{i\phi_1}\sqrt{1-\varepsilon}},\frac{i\sqrt{1-\varepsilon}+\e^{i\phi_2}}{1-i\e^{i\phi_2}\sqrt{1-\varepsilon}},\dots\nonumber\\
&\phantom{{\rm
  diag}(r'_{\varepsilon_1}}\dots,\frac{i\sqrt{1-\varepsilon}+\e^{i\phi_{N_{\phi}}}}{1-i\e^{i\phi_{N_{\phi}}}\sqrt{1-\varepsilon}})\,.\nonumber
\end{align}

Using these expressions together with (\ref{eq:16}) for the voltage
probe, respectively (\ref{eq:12}) for the dephasing probe, we can
express the electrochemical capacitance and the charge relaxation
resistance as a function of the transparency
$\mathcal{T}=1-r^2$ of the current carrying channel, the phases $\phi_1,\dots,\phi_{N_{\phi}}$ and the coupling
strength $\varepsilon$. In order to investigate the crossover from the
coherent to the incoherent regime, we will later on make a specific
physically motivated choice for the energy dependence of the
phases. However, even without specifying the form of the energy-phase
relation, we can already draw some general conclusions by looking at the
incoherent limit $\varepsilon\rightarrow 1$. This we do next after
briefly reviewing the coherent case $\varepsilon=0$.

\subsection{Results and discussion\label{sec:results-discussion}}
\subsubsection{General results at zero temperature\label{sec:general-results-at}}
We first consider the zero temperature limit. In this case voltage and
dephasing probe models are equivalent as shown in
section~\ref{sec:volt-deph-probe}. The capacitance and the resistance
are given by
\begin{equation}\label{eq:23}
C_{\mu}=\frac{-C\sum_{\alpha\beta}A_{\alpha\beta}^1}{-iC-\sum_{\alpha\beta}A_{\alpha\beta}^1}\,,
\end{equation}
and
\begin{equation}\label{eq:22}
R_q = \sum_{\alpha\beta}\left(A_{\alpha\beta}^2-\frac{A^1_{\alpha\phi}A^1_{\phi\beta}}{A^0_{\phi\phi}}\right)\Big/\left(\sum_{\alpha\beta}A_{\alpha\beta}^1\right)^2\,,
\end{equation}
where $A^i_{\alpha\beta}\equiv A^i_{\alpha\beta}(E_F)$ are given in
Eq.~(\ref{eq:9}). In the coherent regime ($\varepsilon=0$), we recover
the universal result $R_q=h/(2e^2)$ for the resistance while the
capacitance is given by $C_{\mu}=Ce^2\nu/(C+e^2\nu)$, with the density
of states of the cavity $\nu(E) =1/(2\pi i)S^{\dagger}(E)dS(E)/dE$, where
$S(E)=\lim_{\varepsilon\rightarrow 0}S_{11}(E)$, for $S_{11}$ given in Eq.~(\ref{eq:15}). In the
opposite, fully incoherent regime ($\varepsilon=1$), we find
\begin{equation}\label{eq:21}
C_{\mu}=\frac{C\frac{e^2}{2\pi}\sum_{i=1}^{N_{\phi}}\phi_i^{\prime}}{C+\frac{e^2}{2\pi}\sum_{i=1}^{N_{\phi}}\phi_i^{\prime}}\,,
\end{equation}
which is independent of $\mathcal{T}$, and
\begin{equation}\label{eq:17}
R_q =
\frac{h}{2e^2}\frac{\sum_{i=1}^{N_{\phi}}(\phi_i^{\prime})^2}{\left(\sum_{i=1}^{N_{\phi}}\phi_i^{\prime}\right)^2}+\frac{h}{e^2}\left(\frac{1}{\mathcal{T}}-\frac{\phi_1^{\prime}}{\sum_{i=1}^{N_{\phi}}\phi_i^{\prime}}\right)\,.
\end{equation}
For a single open dephasing channel ($N_{\phi}=1$), Eq.~(\ref{eq:17})
reduces to
\begin{equation}\label{eq:24}
R_q = \frac{h}{2e^2}+\frac{h}{e^2} \frac{1-\mathcal{T}}{\mathcal{T}}\,.
\end{equation}
Thus, as mentioned in the introduction, if only the current carrying channel is dephased, the charge
relaxation resistance is the sum of a constant interface
resistance~\cite{Imry:86,Landauer:87,Sols:99}
$R_c=h/(2e^2)$ and the original Landauer resistance
$R_L=h/e^2(1-\mathcal{T})/\mathcal{T}$ of the QPC.

\subsubsection{Smooth potential approximation\label{sec:smooth-potent-appr}}
In the following, we will assume that the potential in the cavity is
sufficiently smooth so that the energy dependent part of the
accumulated phase is the same for
all channels. Then
$\phi_1^{\prime}\approx\phi_2^{\prime}\approx\dots\phi_{N_{\phi}}^{\prime}\equiv\phi^{\prime}$. Within this approximation, Eqs.~(\ref{eq:21}) and (\ref{eq:17}) reduce to
\begin{equation}\label{eq:26}
C_{\mu}=\frac{C\frac{e^2}{2\pi}N_{\phi}\phi^{\prime}}{C+\frac{e^2}{2\pi}N_{\phi}\phi^{\prime}}\,,
\end{equation}
and
\begin{equation}\label{eq:25}
R_q =\frac{h}{e^2}\frac{1-\mathcal{T}}{\mathcal{T}}+\frac{h}{2e^2}+\frac{h}{2e^2}\frac{N_{\phi}-1}{N_{\phi}}\,.
\end{equation}
Written in this way, this expression for $R_q$ again lends itself to a simple
interpretation. The first term on the righthand side of
Eq.~(\ref{eq:25}), is the original Landauer resistance
of the QPC. The second term $R_c\equiv\frac{h}{2e^2}$ is the interface resistance of the real
reservoir-conductor interface and the third term
$R_{\phi}=\frac{h}{2e^2}\frac{N_{\phi}-1}{N_{\phi}}$ is the resistance contributed to by the
dephasing. In the limit of a very large number of open dephasing channels
($\varepsilon=1,\ N_{\phi}\gg 1$), which
corresponds physically to spatially homogeneous dephasing,
$R_{\phi}\rightarrow R_c$ and so $R_q\rightarrow h/e^2(1/\mathcal{T})$ as well as
$C_{\mu}\rightarrow C$. Thus, in this limit the fictitious probe
contributes half a resistance quantum and the mesoscopic capacitor
behaves like a classical RC circuit with a two terminal resistance in series with the geometrical gate
capacitance.

Next we investigate the crossover from the coherent to the incoherent
regime. For this purpose, we assume that the accumulated phase depends
linearly on energy in the vicinity of the Fermi energy; explicitly we take $\phi_1(E) =\phi_2(E)=\dots=\phi_{N_{\phi}}(E)=2\pi E/\Delta$, where $\Delta$ is the mean level
spacing in the cavity. Then, the fictitious probe is characterized by only two parameters; the number of channels
$N_{\phi}$ and the coupling strength $\epsilon$. Following
Ref.~\onlinecite{Buttiker:88}, the latter can be related to the dephasing
time $\tau_{\phi}$. The scattering amplitudes have poles at the
complex energies $E=E_n-i\Gamma_e/2-i\Gamma_{\phi}/2$,
where $E_n = (2n+1)\Delta/2$ with $n=0,1,\dots$ is a resonant energy and
$\Gamma_e=-(\Delta/\pi)\ln[r]$ and
$\Gamma_{\phi}=-(\Delta/\pi)\ln[\sqrt{1-\varepsilon}]$ are
respectively the elastic and inelastic widths. The dephasing time
$\tau_{\phi}=\hbar/\Gamma_{\phi}$ is
then related to $\varepsilon$ by
$\varepsilon=1-\exp[-h/(\tau_{\phi}\Delta)]$.

\begin{figure}[ht]
{\psfrag{1a}{\color{red}{$1a$}}
 \psfrag{1b}{\color{red}{$1b$}}
 \psfrag{2a}{\color{blue}{$2a$}}
 \psfrag{2b}{\color{blue}{$2b$}}
\psfrag{3}{$3$}
\includegraphics[width=0.45\textwidth]{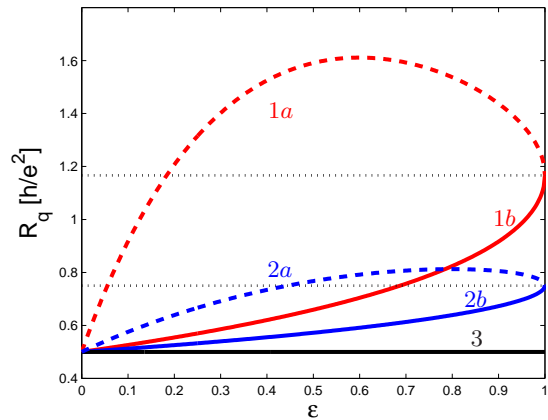}}
\caption{(Color online) $R_q$ as a function of the dephasing strength $\varepsilon$ of a
  single channel probe ($N_{\phi}=1$) at zero temperature, for
  different values of the channel transmission probability
  $\mathcal{T}$. $(1)$: $\mathcal{T}=0.6$, $(2)$: $\mathcal{T}=0.8$ and $(3)$:
  $\mathcal{T}=1$. The dashed curves $(a)$ show the off-resonant case
  $\Delta E\equiv \min_n(E_F-n\Delta)=0.5\Delta$, while
  the full curves $(b)$ show the on-resonant case $\Delta E=0$. The
  horizontal dotted lines represent the value of
  $\frac{h}{2e^2}+\frac{h}{e^2}\frac{1-\mathcal{T}}{\mathcal{T}}$ for
  the different transmission probabilities.\label{fig:3}}
\end{figure}

In Fig.~\ref{fig:3} we show the behavior of $R_q$ as a function of the
dephasing strength $\varepsilon$, if the probe is
weakly coupled so that only one channel ($N_{\phi}=1$)
with transmission probability $\varepsilon$ connects the cavity to the
fictitious reservoir.
We see that if the current carrying channel is
perfectly transmitted through the QPC, i.e. for $\mathcal{T}=1$, the resistance is insensitive to dephasing and is fixed to its
coherent value of half a resistance quantum (curve
$3$). This is reasonable
since for perfect transmission the electronic wavefunction is not
split at the QPC and hence an electron cannot interfere with itself whether or not
it evolves coherently along the edge channel. We emphasize however, that this
simple argument holds only if the dephasing probe couples to a single
channel. If the probe is coupled more strongly, such that it
couples to additional (closed) channels inside the cavity ($N_{\phi}>1$), the ensuing effective
incoherent coupling between channels will affect
$R_q$ in an $\varepsilon$-dependent manner. Turning our
attention back to the single channel case of Fig.~\ref{fig:3}, we see
that as the transparency of the channel is reduced, the charge
relaxation resistance increases with $\varepsilon$. Also evident is
that dephasing affects the resistance non-monotonically in the
off-resonant case (curves a), where the energy of the electron lies between two
Fabry-Perot-like resonances, and monotonically in the on-resonant case (curves
b). This can be related to the fact that dephasing induces both a
decrease of the peak value and a broadening of the density of states
(DOS) in the cavity. On resonance the net result is thus always a monotonous
decrease of the amplitude of the DOS. Off-resonance however, the
amplitude may first increase due to the widening of the closest resonance. Finally, as expected dephasing
is seen to affect the resistance the stronger, the weaker the coupling
to the reservoir is, i.e. the longer an electron dwells inside the
cavity, where it can undergo dephasing.

\subsubsection{Dephasing vs Temperature induced phase averaging\label{sec:deph-vs-temp}}
\begin{figure}[ht]
{\psfrag{a}{$a$}
 \psfrag{bl}{$b$}
 \psfrag{c}{$c$}
 \psfrag{d}{$d$}
 \psfrag{VP}{Voltage Probe}
 \psfrag{DP}{Dephasing Probe}
 \psfrag{h/2e2}{$\frac{h}{2e^2}$}
 \psfrag{h/e2(1/T-1/2)}{$\frac{h}{e^2}\frac{1-\mathcal{T}}{\mathcal{T}}+\frac{h}{2e^2}$}
 \psfrag{h/e2(1/T-1/(2N))}[][l]{$\frac{h}{e^2}\frac{1-\mathcal{T}}{\mathcal{T}}+\frac{h}{2e^2}+\frac{h}{2e^2}\frac{N_{\phi}-1}{N_{\phi}}$}
 \psfrag{N=1, R=0.5, DE=0}{$N_{\phi}=1$, $\mathcal{T}=0.5$, $\Delta E=0$}
 \psfrag{N=2, R=0.5, DE=0}{$N_{\phi}=2$, $\mathcal{T}=0.5$, $\Delta E=0$}
\includegraphics[width=0.45\textwidth]{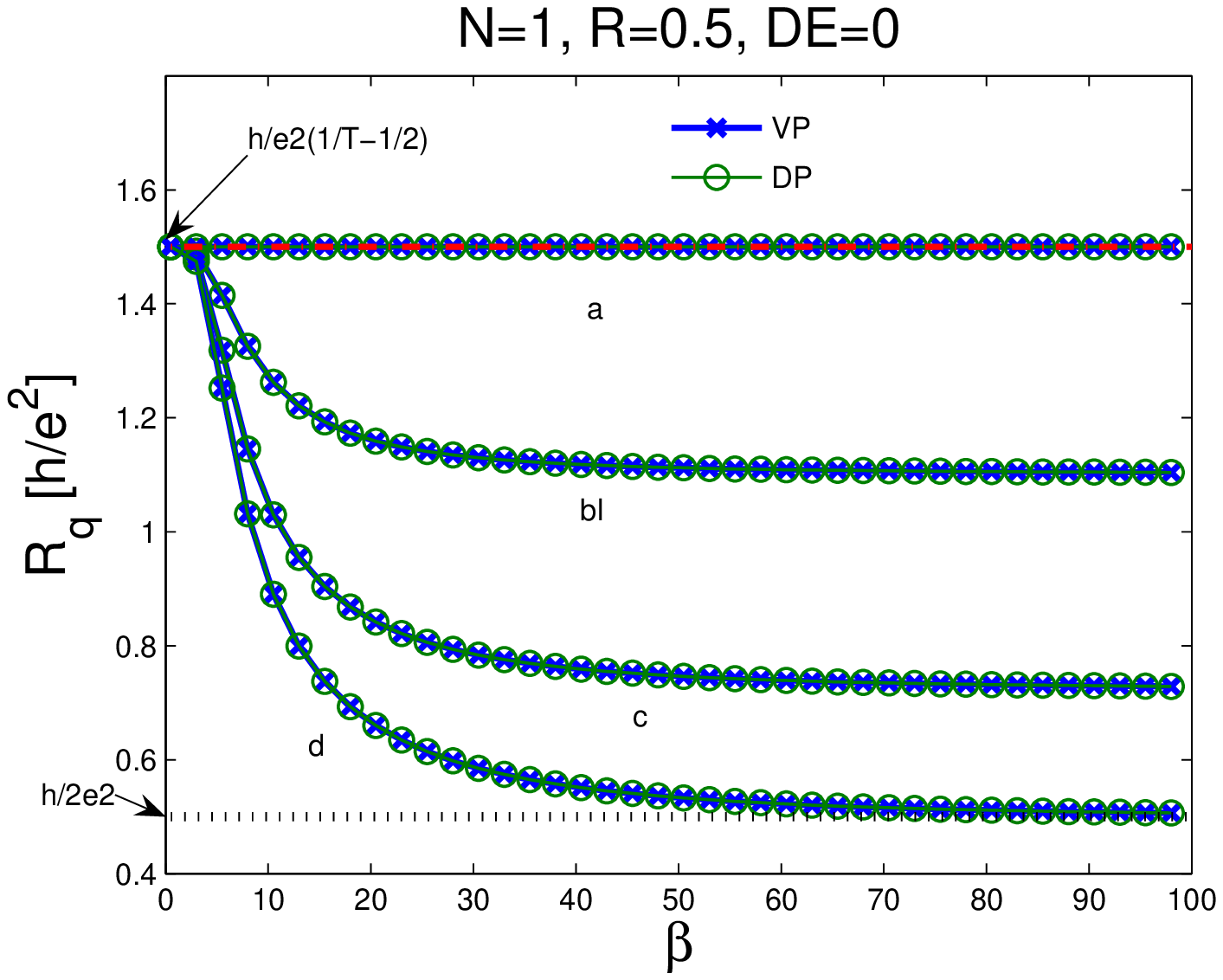}\\\vfill
\includegraphics[width=0.45\textwidth]{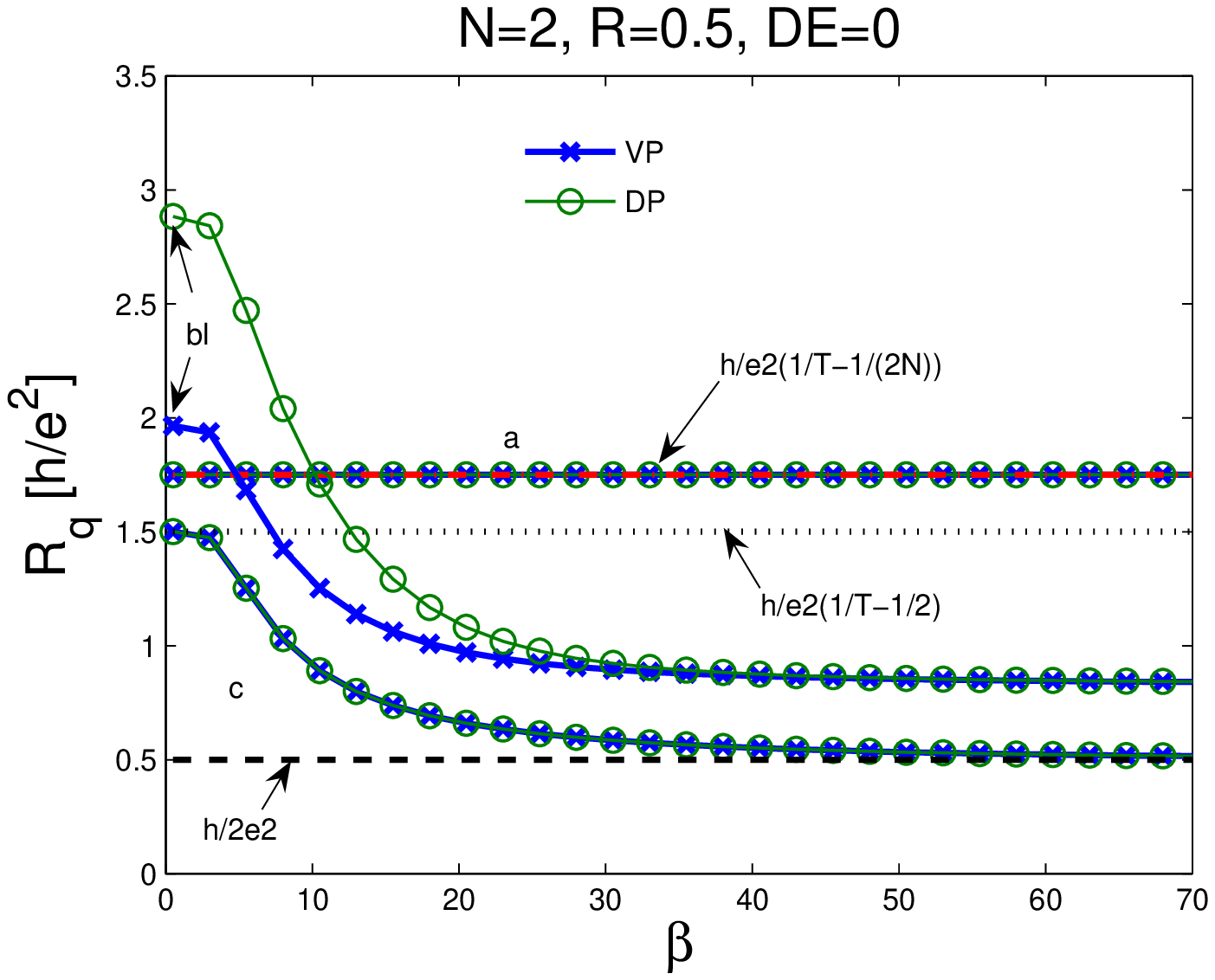}}
\caption{(Color online) {\it Upper panel}: $R_q$ as a function of the inverse temperature $\beta$ for
  $N_{\phi}=1$ and dephasing strengths: ($a$) $\varepsilon=1$,
  ($b$) $\varepsilon=0.9$, ($c$) $\varepsilon=0.5$ and ($d$) $\varepsilon=0$. The
  dashed-dotted red line gives the value
  $h/e^2(1-\mathcal{T})/\mathcal{T}+h/(2e^2)$ and the
    dotted line the value $h/(2e^2)$. As discussed
  in the text, dephasing and voltage probes are
  indistinguishable in this case. {\it Lower panel}: $R_q$ as a function of the inverse temperature $\beta$ for
  $N_{\phi}=2$ and dephasing strengths: ($a$) $\varepsilon=1$,
  ($b$) $\varepsilon=0.7$ and ($c$) $\varepsilon=0$. The
  dashed-dotted red line gives the value
  $h/e^2(1-\mathcal{T})/\mathcal{T}+\frac{h}{2e^2}+\frac{h}{2e^2}(N_{\phi}-1)/(N_{\phi})$, the
    dotted line the value
    $h/e^2(1-\mathcal{T})/\mathcal{T}+h/(2e^2)]$ and
    the dashed line the value of $h/(2e^2)$. As discussed
  in the text, dephasing and voltage probes differ for
  finite dephasing if $\varepsilon\not=1$, (curves $b$). $\beta$ is
  given in units of the inverse level spacing $\Delta^{-1}$. We show here the resonant
    cases $\Delta E\equiv \min_n(E_F-n\Delta)=0$; the off-resonant behavior is similar.\label{fig:6}}
\end{figure}

It is instructive to compare the results obtained in the previous
section in the incoherent limit $\varepsilon=1$ at zero temperature with finite temperature
effects in the coherent regime $\varepsilon=0$. At finite temperature and
for a perfectly coherent single channel system, the charge relaxation
resistance is given by~\cite{Buttiker:93b}
\begin{equation}\label{eq:1}
R_q = \frac{h}{2e^2}\frac{\int dE(-f'(E))\nu(E)^2}{\left(\int dE(-f'(E))\nu(E)\right)^2}\,,
\end{equation}
where $\nu(E)$ is the density of states of the channel which was
defined above and is here explicitly given by~\cite{Gabelli:06}
$\nu(E)=(1/\Delta)\left[(1-r^2)/(1+r^2+2r\cos(2\pi E/\Delta))\right]$. At low
temperature $k_BT\ll\Delta$, an expansion around the Fermi energy
yields
$R_q=\frac{h}{2e^2}\left(1+\frac{\pi^2}{3\beta^2}\left(\ln'[\nu(E_F)]\right)^2\right)$
with $\beta=1/(k_BT)$. Finite temperature effects thus arise at order $T^2$ and lead to the
appearance of pairs of
peaks in the resistance as a function of Fermi energy around each
resonance, where the square of the derivative of the density of states
is maximal~\cite{Nigg:06} (see thin red curve in Fig.~\ref{fig:5}, top). This behavior has indeed been
observed experimentally~\cite{GabelliThesis:06} in the weakly coupled
regime, where $\Delta\gg k_BT\gg\mathcal{T}\Delta$. At very high temperature $k_BT\gg\Delta$, the
integrals in (\ref{eq:1}) may be evaluated asymptotically as shown in the
appendix and we obtain
$R_q=h/(2e^2)+h/e^2(1-\mathcal{T})/\mathcal{T}$.
Thus phase averaging in the high temperature coherent regime ($\varepsilon=0$)
is equivalent to dephasing via a fictitious probe with a single
open channel ($\varepsilon=1$) at zero temperature (see Eq.~(\ref{eq:24})). This fact and the
crossover from the low to the high temperature regime are illustrated in the upper panel of
Fig.~\ref{fig:6}. There we show the charge relaxation resistance as a
function of the inverse temperature $\beta$ for different
dephasing strengths $\varepsilon$ for $N_{\phi}=1$. For complete dephasing
(curve a), $R_q$ is temperature independent and given by
Eq.~(\ref{eq:17}) with $N_{\phi}=1$. Interestingly, we find that for a
single channel probe, voltage and dephasing probes equally affect the
resistance even at finite temperature. Technically this is due to the fact that for a linear
energy-phase relation such as assumed in this work, the energy dependent parts of each factor in the
integrands of Eq.~(\ref{eq:18}) below are identical.

{\it Dephasing vs Voltage probe}. At the end of the last paragraph, we
concluded that dephasing and voltage probes are indistinguishable for
a single channel probe as long as the accumulated phase is linear in energy. When $N_{\phi}\geq 2$, the two
dephasing models differ at finite
temperature. Introducing the
{\em emittances}~\cite{Pretre:96} $\mathcal{N}_{\alpha}^{{\rm
    em}}\equiv 1/(2\pi
  i)\sum_{\beta}\tr[S_{\alpha\beta}^{\dagger}S^{\prime}_{\alpha\beta}]=1/(ie^2)\sum_{\beta}A^1_{\alpha\beta}$, which represent the DOS of carriers emitted into probe $\alpha$ and the {\em injectances} $\mathcal{N}_{\beta}^{{\rm in}}\equiv1/(2\pi
 i)\sum_{\alpha}\tr[S_{\alpha\beta}^{\dagger}S^{\prime}_{\alpha\beta}]=1/(ie^2)\sum_{\alpha}A^1_{\alpha\beta}$ representing the DOS of carriers injected from probe $\beta$, we may write the difference of resistance between the two models $\Delta R_q\equiv R_q^{{\rm DP}}-R_q^{{\rm VP}}$ as
\begin{equation}\label{eq:18}
\Delta R_q = \frac{h}{e^2}\frac{\frac{\int dE
    f'\mathcal{N}_{\phi}^{{\rm em}}\int dEf'\mathcal{N}_{\phi}^{{\rm in}}}{\int dEf'\tr[\openone_{\phi}-S^{\dagger}_{\phi\phi}S_{\phi\phi}]}-\int dE
    f'\frac{\mathcal{N}_{\phi}^{{\rm em}}\mathcal{N}_{\phi}^{{\rm in}}}{\tr[\openone_{\phi}-S^{\dagger}_{\phi\phi}S_{\phi\phi}]}}{\left(\int
    dEf'\mathcal{N}\right)^2}\,,
\end{equation}
where
$\mathcal{N}=\sum_{\alpha}\mathcal{N}_{\alpha}^{{\rm
    in}}=\sum_{\alpha}\mathcal{N}_{\alpha}^{{\rm em}}$
is the total DOS and for compactness we have suppressed all the energy
arguments. As illustrated in
the lower panel of Fig.~\ref{fig:6}, we find somewhat
counter-intuitively, that the
resistance is larger in the presence of a dephasing probe than in the
presence of a voltage probe. Indeed, one would have expected that
since the voltage probe is dissipative and the dephasing probe is not,
the former should lead to a larger resistance than the
latter. This intuition fails when applied to $R_q$. Finally, for complete dephasing (curve $a)$), the two models coincide
again. This is due to the fact that for $\varepsilon=1$ the
coefficients given in (\ref{eq:9}) become energy independent as a
consequence of the linear energy-phase relation.

\subsubsection{Comparison with experiment\label{sec:comp-with-exper}}
\begin{figure}[ht]
{\includegraphics[width=0.45\textwidth]{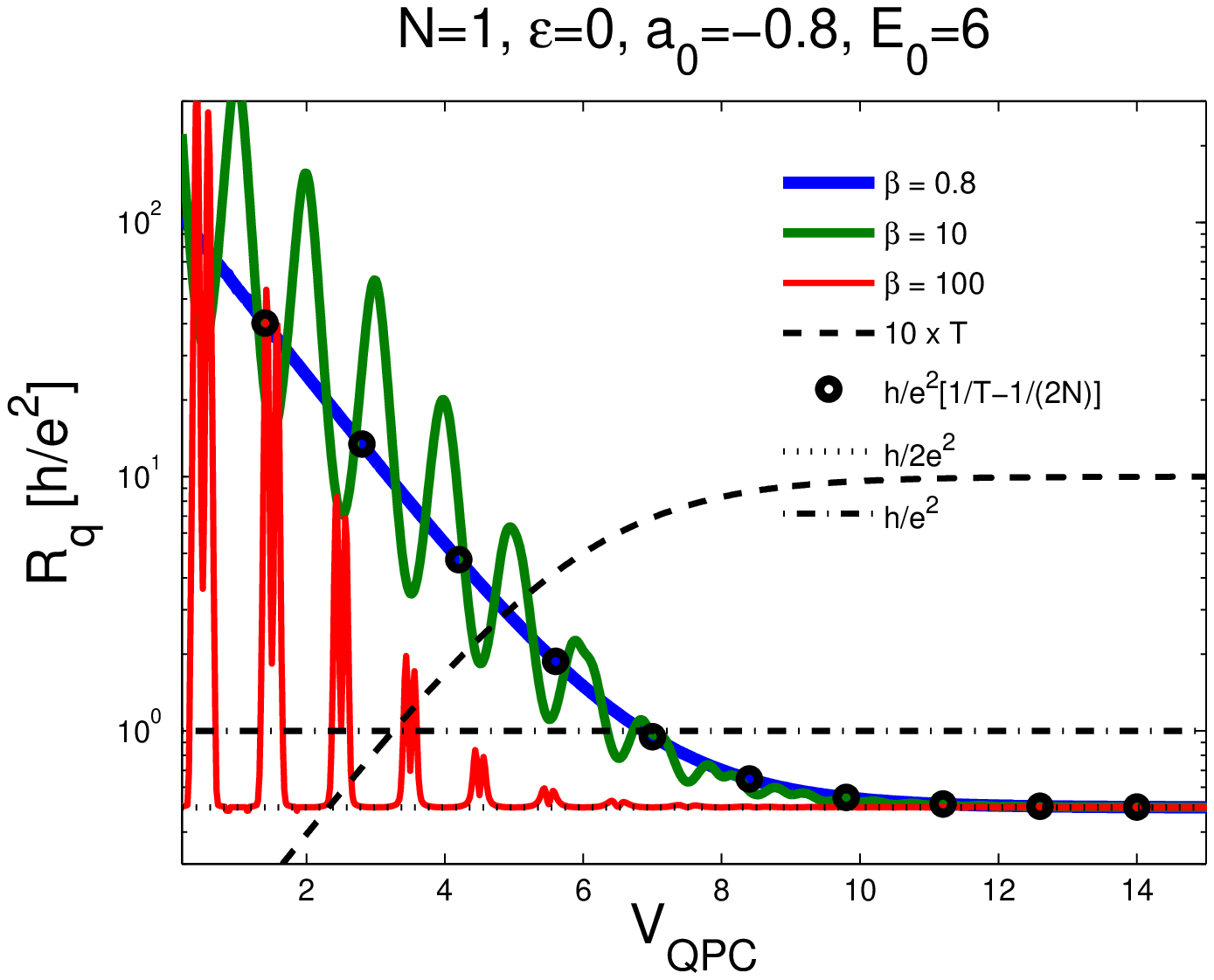}}\\\vfill
{\includegraphics[width=0.45\textwidth]{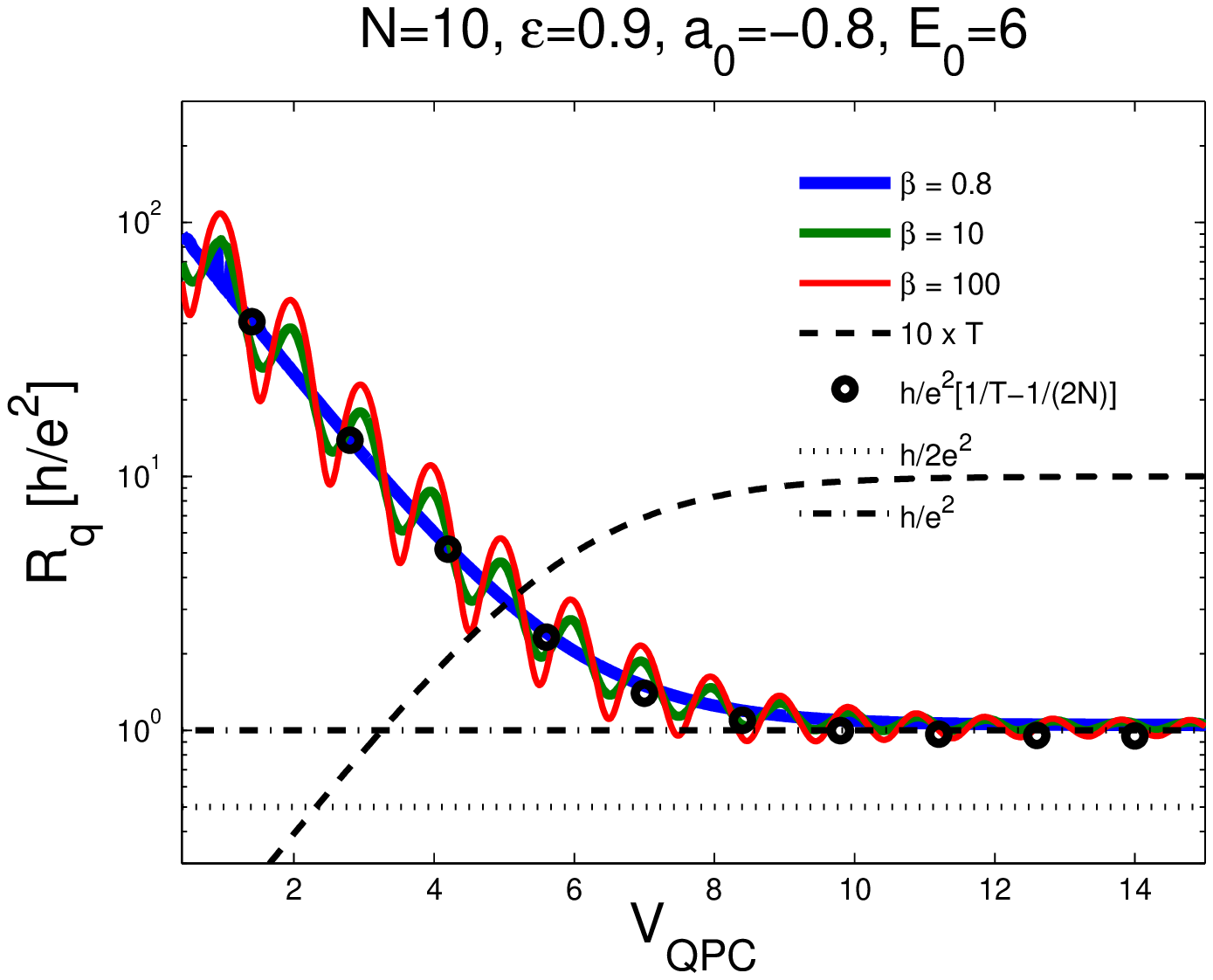}}
\caption{(Color online) $R_q$ as a function of $V_{QPC}$ for different temperatures for a coherent
  system $\varepsilon=0$ (upper panel) and for a strongly incoherent system $\varepsilon=0.9$ and $N_{\phi}=10$ (lower
  panel). The inverse
  temperature $\beta$ is given in units of the inverse level
  spacing $\Delta^{-1}$. \label{fig:5}}
\end{figure}

Comparison with the experiment~\cite{Gabelli:06}, leads us to some
important conclusions. In this experiment, the real and imaginary parts of the AC conductance
(\ref{eq:2}) were measured at $mK$ temperatures $k_BT\ll\Delta$ while varying the transmission of the QPC,
giving access to the charge relaxation resistance over a wide range of
channel transparencies. In the highly
transmissive regime the quantization of the charge relaxation
resistance of a single channel mesoscopic capacitor could thereby be
verified. As the coupling to the lead was reduced an oscillating
increase in resistance was measured and excellent agreement with a
theoretical model including only temperature broadening effects was obtained in the
regime $\Delta\gg k_BT\gg\mathcal{T}\Delta$. For higher
temperatures $T\sim 4K>\Delta/k_B$ the resistance was
found to approach $\frac{h}{e^2}$ for a single perfectly open channel~\cite{GabelliThesis:06}
indicating that in this regime, the cavity truly acts like an
additional reservoir. Indeed, according to
our discussion in section \ref{sec:deph-vs-temp}, pure phase
averaging due to temperature broadening would instead lead to
$R_q\xrightarrow{k_BT\gg\Delta}\frac{h}{2e^2}$ for
$\mathcal{T}=1$. Thus the observed value of the resistance hints at
the presence of a spatially
homogenous dephasing mechanism effective at high
temperatures, which is suppressed at low temperatures. One such mechanism is the thermally activated tunneling from the current carrying edge channel to nearby localized states, which together act as a many channel voltage (dephasing) probe depending on
the energetics of the scattering process. In Fig.~\ref{fig:5}, we show the charge
relaxation resistance as a function of the QPC voltage. The dependence
of the transmission probability on $V_{QPC}$ is modeled assuming that
the constriction is well described by a saddle-like
potential~\cite{Buttiker:90,Gabelli:06} $\mathcal{T}=[1+\exp\{-a_0(V_{QPC}-E_0)\}]^{-1}$ and a change in $V_{QPC}$ is assumed to
induce a proportional shift in the electrostatic potential of the
QD. In the upper panel we show the coherent case $\varepsilon=0$ for
three different temperatures $k_BT/\Delta = 0.8,\ 0.1$ and $0.01$. At low temperature and small transmission we recognize
the resistance oscillations discussed in section~\ref{sec:deph-vs-temp}. As the temperature
is increased, $R_q$ goes towards $h/e^2(1-\mathcal{T})/\mathcal{T} +h/(2e^2)$. This
is to be contrasted with the situation shown in the lower panel where
we display the incoherent case for a voltage probe  with $\varepsilon=0.9$,
$N_{\phi}=10$ and the same set of temperatures. For an open constriction
($\mathcal{T}\approx 1$), $R_q$ is now close to $h/e^2$ in the high
temperature regime in agreement with the experimental observation.

\section{Conclusion\label{sec:conclusion}}
In this work, we investigate the effect of decoherence on the dynamic
electron transport in a mesoscopic capacitor. Extending the voltage
and dephasing probe models to the AC regime, we calculate the charge
relaxation resistance and the electrochemical capacitance, which
together determine the RC-time of the system. Dephasing breaks the
universality of the single channel, zero temperature charge relaxation
resistance and introduces a dependency on the transparency of the QPC. We find that
complete {\em intra-channel} relaxation alone is not sufficient to recover the two
terminal resistance formula but rather yields a resistance
which is the sum of the original Landauer formula and the interface
resistance to the reservoir. This is also the resistance obtained in
the high temperature limit of the coherent single channel system. Only in the
presence of perfect {\em inter-channel} relaxation with a large number of channels, does the QD
act as an additional reservoir and we recover the classically expected
two terminal resistance.

\begin{acknowledgements}
We thank Heidi F{\"o}rster and Mikhail Polianski for helpful comments on
the manuscript. This work was supported by the Swiss NSF, the STREP project SUBTLE and the Swiss National Center of
Competence in Research MaNEP.
\end{acknowledgements}
\appendix
\section{High temperature regime
  integrals\label{sec:high-temp-regime}}
In this appendix we compute the integrals appearing in
the high temperature limit of Eq.~(\ref{eq:1}). Asymptotically we
have
\begin{equation}\label{eq:10}
\lim_{\beta\rightarrow 0}R_q = \frac{h}{2e^2}\frac{I_2}{(I_1)^2}\,,
\end{equation}
where 
\begin{equation}
I_1=\int_0^{\Delta}dE\nu(E),\quad\text{and}\quad I_2=\Delta\int_0^{\Delta}dE\nu(E)^2\,
\end{equation}
with $\nu(E)=\frac{1}{\Delta}\frac{1-r^2}{1+2r\cos\frac{2\pi
    E}{\Delta}+r^2}$.
Following a change of variables $x=\frac{2\pi E}{\Delta}$ we get
\begin{equation}
I_1=\frac{1}{2\pi}\int_0^{2\pi}dx\frac{1-r^2}{1+2r\cos
  x+r^2}\,,
\end{equation}
and
\begin{equation}
I_2=\frac{1}{2\pi}\int_0^{2\pi}dx\frac{(1-r^2)^2}{(1+2r\cos x+r^2)^2}\,.
\end{equation}
A simple way of computing these integrals is to use the (Poisson
kernel) identity
\begin{equation}
\frac{1-r^2}{1+2r\cos x+r^2}=\sum_{k=-\infty}^{\infty}(-r)^{|k|}\e^{ikx}\,,
\end{equation}
which can easily be verified by splitting the sum as
$\sum_{k=-\infty}^{\infty}x^k=\sum_{k=0}^{\infty}x^k+\sum_{k=-\infty}^{0}x^k-1$
and utilizing the fact that for $|r|<1$, the geometric series
converge.
Integrating the sum in $I_1$ term by term and using the identity
$\int_0^{2\pi}\e^{ikx}dx=2\pi\delta_{k0}$ for $k\in\mathbb{Z}$, we
immediately find $I_1 = 1$. Similarly we have
\begin{align}
I_2 &=
\frac{1}{2\pi}\int_0^{2\pi}dx\left(\sum_{k=-\infty}^{\infty}(-r)^{|k|}\e^{ikx}\right)^2\nonumber\\
&=\frac{1}{2\pi}\sum_{k,k'=-\infty}^{\infty}(-r)^{|k|+|k'|}\int_0^{2\pi}dx\e^{i(k+k')x}\nonumber\nonumber\\
&=\sum_{k,k'=-\infty}^{\infty}(-r)^{|k|+|k'|}\delta_{k+k',0}=\frac{1}{\Delta}\sum_{k=-\infty}^{\infty}(-r)^{2|k|}\nonumber\\
&=\sum_{k=0}^{\infty}(-r)^{2k}+\sum_{k=-\infty}^{0}(-r)^{-2k}-1\nonumber\\
&=\frac{2}{1-r^2}-1=\frac{1+r^2}{1-r^2}\,.
\end{align}
Substituting back into (\ref{eq:10}) yields the desired result.

\bibliography{biblio}
\end{document}